\documentclass[manuscript]{aastex}
\usepackage{graphicx}
\usepackage{apjfonts}
\usepackage{amsmath}
\usepackage{natbib}
\usepackage{xcolor}

\shorttitle{LAXPC timing study of 4U 0115$+$63}
\shortauthors{Authors et al.}

\begin{document}
\title{LAXPC / AstroSat Study of $\sim$ 1 and $\sim$ 2 mHz Quasi-periodic Oscillations in the Be/X-ray Binary 4U~0115+63 During its 2015 Outburst}


\author{Jayashree Roy $^{1,2}$, P. C. Agrawal $^{1}$, N. K. Iyer $^{2}$, D. Bhattacharya $^{2}$, J. S. Yadav $^{3}$, H. M. Antia $^{3}$, J. V. Chauhan $^{3}$, M. Choudhury $^{1}$, D. K. Dedhia $^{3}$, T. Katoch $^{3}$, P. Madhavani $^{3}$, R.K. Manchanda $^{4}$, R. Misra $^{2}$, M. Pahari $^{2}$, B. Paul $^{5}$, P. Shah$^{3}$ \\}
\affil{$^{1}$UM-DAE Centre for Excellence in Basic Sciences, University of Mumbai, Vidyanagari Campus, Kalina, Santacruz (East), Mumbai, Maharashtra 400098, India.\\
$^{2}$Inter-University Center for Astronomy and Astrophysics, Post Bag 4, Pune, Maharashtra 411007, India. \\
$^{3}$Tata Institute of Fundamental Research, Homi Bhabha Road, Navy Nagar, Colaba, Mumbai, Maharashtra 400005, India.\\
$^{4}$Department of Physics, Mumbai University, Kalina, Santacruz East, Mumbai, Maharashtra 400098, India.\\
$^{5}$Raman Research Institute, Sadashivnagar, Bangalore, Karnataka 560080, India.}
\email{jayashree@iucaa.in}



\begin{abstract}
The Be X-ray Binary 4U 0115+63 was observed by Large Area X-ray Proportional Counter (LAXPC) instrument on AstroSat on 2015 October 24 during the peak of a giant Type II outburst. Prominent intensity oscillations at $\sim$ 1 and $\sim$ 2 mHz frequency were detected during the outburst. Nuclear Spectroscopic Telescope Array (NuSTAR) observations made during the same outburst also show mHz quasi periodic oscillations (QPOs). Details of the oscillations and their characteristics deduced from LAXPC/AstroSat and NuSTAR observations are reported in this paper. Analysis of the archival Rossi X-ray Timing Explorer (RXTE) / Proportional Counter Array (PCA) data during 2001-11 also show presence of mHz QPOs during some of the outbursts and details of these QPOs are also reported. Possible models to explain the origin of the mHz oscillations are examined. Similar QPOs, albeit at higher frequencies, have been reported from other neutron star and black hole sources and both may have a common origin.
Current models to explain the instability in the inner accretion disk causing the intense oscillations are discussed.
\end{abstract}


\keywords{X-rays: binaries - pulsars: individual 4U 0115+634}
\section{Introduction}
The high mass X-ray binary (HMXB) 4U 0115+63, originally discovered by the UHURU satellite \citep{1972ApJ...178..281G, 1978ApJS...38..357F}, has been studied extensively and found to be a Be system with an orbital period of 24.3 days \citep{1978ApJ...224L...1R}. The X-ray source was found to be pulsating with 3.61 sec period by \citet{1978IAUC.3163....1C} from SAS-3 observations. The pulse has a double-peak profile below 20 keV which changes to a single peak shape above 20 keV. During most of its orbit the X-ray binary is in its quiescent state with X-ray luminosity (L$_{x}$) $\le$ 10$^{34}$ erg/s \citep{2001ApJ...561..924C}. When the X-ray source is in active state, it exhibits two types of outbursts (i) Type I bursts which occur regularly once in every orbital cycle near the periastron passage of the neutron star with L$_{x}$ $\sim$ 10$^{36}$-10$^{37}$ ergs/s \citep{1986ApJ...308..669S} and (ii) Type II bursts that occur irregularly and have much higher intensity with peak intensity in the range of $\sim$0.1 - 1 Crab (L$_{x}$ $\ge$ 10$^{37}$ ergs/s) \citep{2013AstL...39..375B}. A detailed study of the outbursts from 4U 0115+63 carried out by \citet{2013AstL...39..375B} and \citet{2015ATel.8231....1B} revealed occurrence of 16 Type II bursts from this source in 1969 - 2015 period. There is a hint of a $\sim$ 3-5 years recurrence period of Type II burst. This has been explained as arising from the loss and reformation of a circumstellar disk around the V=15.5 magnitude Be companion star V635 \citep{2001A&A...369..117N}. Distance to the source has been estimated to be 7 kpc \citep{2001A&A...369..108N}. During the Type II outbursts, the source has been found to spin-up. A spin-up rate of $\dot{P}$=(-7.24$\pm$0.03)$\times$10$^{-6}$ s/d was measured by \citet{2012MNRAS.423.2854L} during the 2008 outburst.

A cyclotron line in the spectrum of 4U 0115+63 was first reported by \citet{1979Natur.282..240W} at about 20 keV. Subsequent studies revealed that this was the first harmonic of a Cyclotron Resonant Scattering Feature (CRSF) in absorption at about 11 keV.

Five CRSFs, were first detected at $\sim$ 11.2, 22.9, 32.6, 40.8, and 53 keV by \cite{2009A&A...498..825F} using BeppoSAX data. Using different continuum models \citet{2013AstL...39..375B} confirmed the presence of a fundamental cyclotron line and 4 higher harmonics at $\sim$ 11, 24, 35.6, 48.8, and 60.7 keV from INTEGRAL data during the 2011 outburst of the source. Using 11 keV as the fundamental line energy, the magnetic field of the neutron star in the region, where CRSFs originate, can be inferred to be $\sim$ 10$^{12}$ Gauss.

Besides the regular 3.61 seconds pulsations, 4U 0115+63 also shows QPOs and their harmonics in the power density spectra (PDS) of its light curves. \citet{1989ESASP.296..617S} detected a broad QPO peak at $\sim$ 62 mHz during the 1978 outburst. A low frequency QPO at $\sim$ 2 mHz was detected during March 1999 outburst \citep{1999ApJ...521L..49H} while QPOs with frequency varying in $\sim$ 27 - 46 mHz range were reported from RXTE/PCA observations during 1999, 2004 and 2008 outbursts \citep{2013MNRAS.434.2458D}. In the most recent type II outburst on 2015 October 22, \citet{2015ATel.8231....1B} detected a $\sim$ 600 sec QPO in the NuSTAR and Swift data which they interpreted as the $\sim$ 2 mHz QPO reported earlier by \citet{1999ApJ...521L..49H}.

In this paper we confirm the $\sim$ 2 mHz QPO observed by RXTE and NuSTAR, and find a $\sim$ 1 mHz intensity oscillations from LAXPC observations of 4U 0115+63 on 2015 October 24, during a giant outburst. 

After the introduction we present a brief description of the LAXPC instrument, details of observation and data analysis methodology. Results from the timing analysis of LAXPC data are highlighted. We have also analyzed data from two NuSTAR observations of 2015 October 22 and October 30, during the same outburst and present results from this analysis for comparison. We have also analysed RXTE/PCA observations of 4U 0115+63 during different outbursts between 2001 and 2011, to search low frequency mHz QPOs not reported earlier from the source and these results are also discussed. Finally we discuss the possible origin of the QPOs and examine the most plausible model.

\section{LAXPC Instrument, Observation \& Data Reduction}

LAXPC instrument on AstroSat consists of 3 identical collimated detectors, having 5 anode layer geometry with 15 cm deep X-ray detection volume providing an effective area of about 4500 cm$^{2}$ at 5 keV, 6000 cm$^{2}$ at 10 keV  and about 5600 cm$^{2}$ at $\sim$ 40 keV. LAXPCs are filled with Xenon-Methane (90\% :10\%) mixture at 1520 torr and have a field of view of 0.9$^{\circ}$ $\times$ 0.9$^{\circ}$ . A description of the AstroSat observatory and its instrument is provided in \citet{2006AdSpR..38.2989A} and \citet{2014SPIE.9144E..1SS}. More detailed description of the characteristics of the LAXPC instrument can be found in \citet{2016SPIE.9905E..1DY}, \citet{2017JApA...38...30A}, \citet{2016ExA....42..249R} and calibration details in \citet{2017ApJS..231...10A}. High time resolution studies to investigate rapid intensity variations and broad band spectral measurements are principal objectives of this instrument. For this purpose the arrival time of every detected X-ray photon is tagged to an accuracy of 10 $\mu$sec, its energy determined by a 1024 channel Pulse Height Analyser and the anode layer in which it interacted, are recorded. After the launch of AstroSat on 2015 September 28, the LAXPC instrument was turned on and became operational on 2015 October 19.
On being alerted that 4U 0115+63 is undergoing a major outburst \citep{2015ATel.8231....1B}, the AstroSat was pointed at this pulsar and observations were made with LAXPC instrument on 2015 October 24 from 09:17:02 UT to 16:08:46 UT covering 4 orbits. Background data for LAXPC were acquired during 2015 October 22 15:31:11 to 2015 October 23 01:48:59 UT by pointing at a source-free region. We have used background data from 2015 October 22 for obtaining the background subtracted light curve of the source. Complete details of the source and background observations are summarized in Table \ref{table1.tab}. 

\begin{table}
\small{
\caption{Log of observations of 4U 0115+63 and Background for LAXPC and NuSTAR.} 
\label{table1.tab}
\begin{tabular}{ccccc}
\hline\hline
Observations&Instrument & Time of Observation &MJD   \\
                   &&        UT     &                      \\
            &&       (yyyy-mm-dd hr:min:sec)   &                 \\
\hline\\
4U 0115+63&NuSTAR&2015-10-22 12:16:08-2015-10-23 04:06:08&57317.5-57318.1\\
          &LAXPC&2015-10-24 09:17:02-2015-10-24 16:08:46&57319.3-57319.6\\
	  &NuSTAR&2015-10-30 13:46:08-2015-10-31 01:16:08&57325.6-57326.0\\
Background&LAXPC&2015-10-22 22:19:24.5-2015-10-23 02:01:42.03&57317.9-57318.0\\
\hline
\end{tabular}
}
\end{table}

Data reduction has been done using LAXPC data reduction pipeline software AS1LAXPCLevel2DataPipeline version 1.0. Conversion of Level-1 raw data file to Level-2 data has been performed by "lxplevel2datapipeline". Data are independently analyzed for each of the 3 LAXPC units. Each LAXPC detector has 7 outputs from 5 anode layers, 2 outputs each from 
anode layers 1 and 2 and 1 output from each of the anode layers 3, 4 and 5. Level-2 data contain (i) light curve in broad band counting mode (modeBB) and (ii) event mode data (modeEA) with information about arrival time, pulse height and layer of origin of each detected X-ray and (iii) housekeeping data and parameter files are stored in mkf file. The light curves were extracted from event mode data using the tool "lxplc" for the source and the background. We have used average background counts in different energy bands for deducing background subtracted source light curves. 
The energy spectrum, response matrix and the background spectrum were extracted using laxpc software ``LaxpcSoft\footnote{http://astrosat-ssc.iucaa.in/?q=data\_and\_analysis}'' having single routine to extract source spectra, light curves and background spectra. The data were analyzed using HEASOFT 6.19\footnote{http://heasarc.gsfc.nasa.gov/docs/software/lheasoft/}. HEASOFT consists of (mainly) FTOOLS for general data extraction and analysis, XRONOS \citep{1992EMIS..59...59S} for the timing analysis and XSPEC package \citep{1996ASPC..101...17A} for the spectral analysis. Table \ref{table2.tab} lists details of the observations and count rates of 4U 0115+63 included in this analysis.

\subsection{NuSTAR and RXTE /PCA Observations and Data reduction}
The NuSTAR data consist of the two focal plane modules A and B (FPMA and FPMB) where each module has a field-of-view (FoV) of 13$'$$\times$13$'$ \citet{2013ApJ...770..103H}. NuSTAR data were obtained from HEASARC data archive for 2015 October 22 and 2015 October 30. Raw event lists from observation ID (ObsID: 90102016002 and 90102016004) were reprocessed with nupipeline, which is part of the NuSTAR Data Analysis Software (NuSTARDAS v1.6.0), while employing the most recent calibration database files available at the time (CALDB: 2015 May 26). Cleaned images were generated for each module in the 3-79 keV energy band. Source and background light curves and energy spectra, have been generated from the cleaned event lists of each of the two focal plane module using a 30"-radius circle around 4U 0115+63. Light curves extracted from FPMA and FPMB were merged using ftools task "fmerge" for timing analysis. The FPMA and FPMB energy spectra are merged using ftools task "addspec" for spectral analysis. Corresponding arf and rmf files are also merged using ftools task "addarf" and "addrmf". 

We have also analysed RXTE/PCA observations of 4U 0115+63 during 2001-2011 outbursts to search low frequency QPO. The PCA consists of five identical Proportional Counter Units (PCUs) sensitive in the 2-60 keV range. The data selection criteria were used to remove the South Atlantic Anomaly (SAA) passage times and the stretches of the observations for which the Earth elevation was $>$10${^\circ}$ and the pointing offset was $>$0${^\circ}$.02. Light curves are extracted from standard 1 mode RXTE/PCA data using ftools task "saextrct" from all the PCUs which were on during the observations.  The three PCUs were on during 2004 observation, while two PCUs were on during the 2008 observations respectively. Moreover, during another observation made in 2011, only one PCU was found to be on in the first stretch of the data and two PCUs were on during the second. We have used data only from PCU 2 for the spectral analysis as this pcu was consistently on during all the observations during the different outbursts. The X-ray spectra of RXTE/PCA were extracted using ftools task "saextrct" using standard 2 mode data with 16s binning. Bright background model pca\_bkgd\_cmbrightvle\_eMv20020201.mdl\footnote{https://heasarc.gsfc.nasa.gov/docs/xte/recipes/pcabackest.html} was used to estimate the background for the spectral analysis. 

\begin{figure}
\includegraphics[scale=0.36,angle=-90]{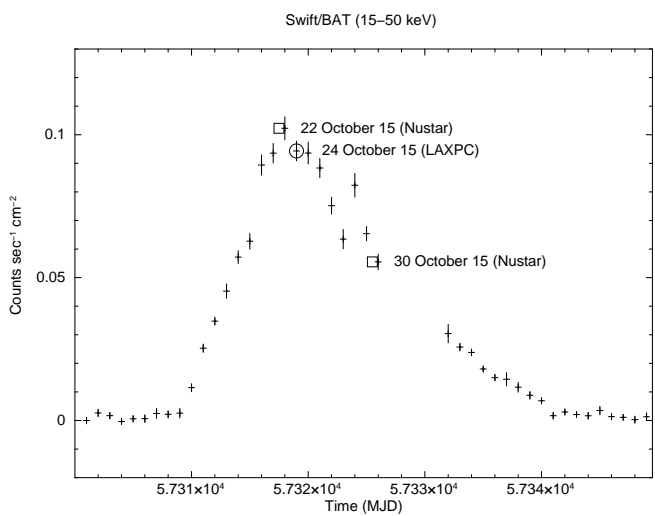}
\caption{Swift/BAT hard X-ray transient monitor light curve in 15-50 keV energy band during 2015 observation of 4U 0115+63. LAXPC observation is indicated with an open circle and NuSTAR observations are shown by open square.}
\label{fig_1}
\end{figure}
\begin{figure}

\includegraphics[scale=0.5,angle=-90]{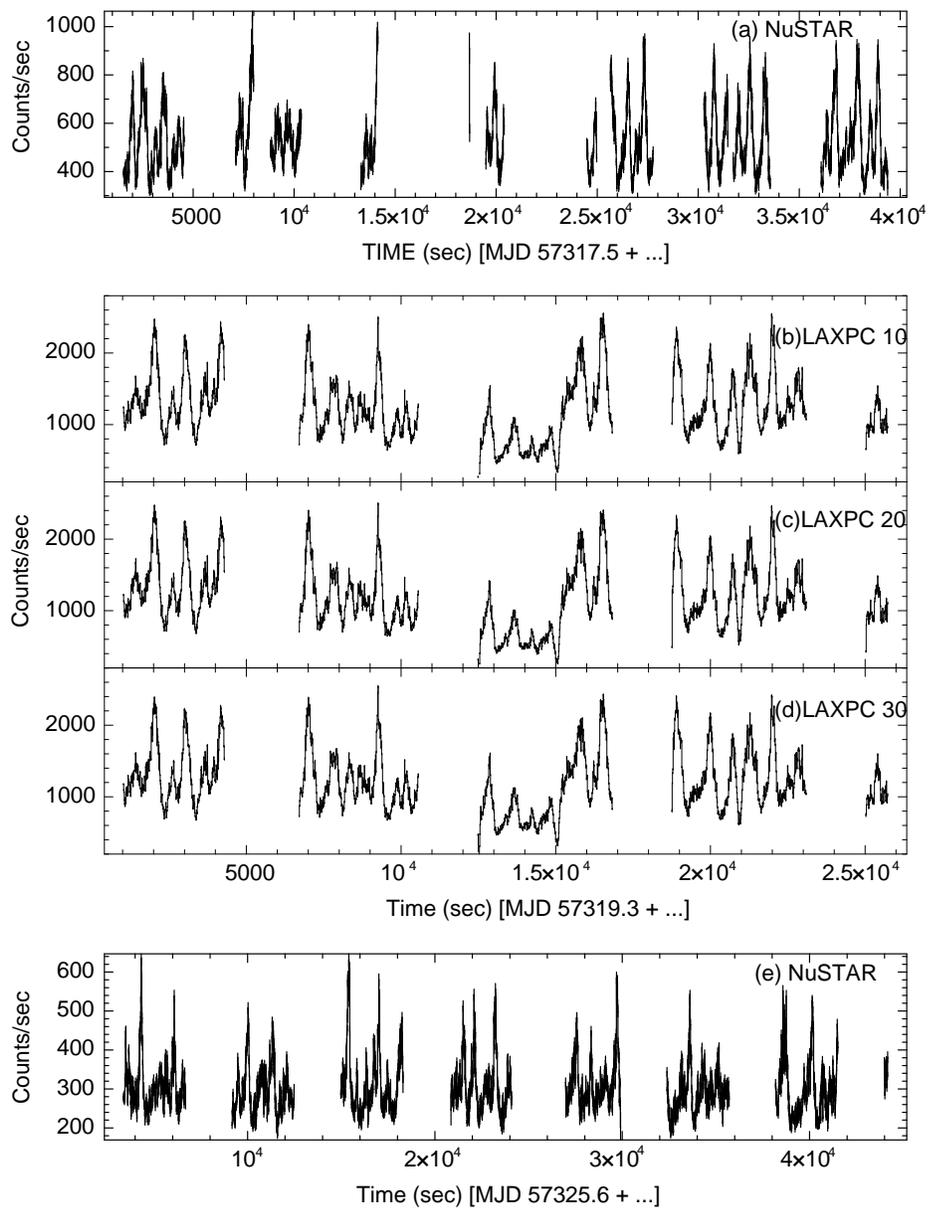}
\caption{Background subtracted light curves in 3-79 keV energy band obtained from the NuSTAR observations are shown in panels, (a) MJD 57317.5 and (e) MJD 57325.6. Background subtracted 3-80 keV lightcurves for the 3 LAXPCs units (LAXPC 10, 20 and 30) on MJD 57319.3 are shown in panels (b), (c) and (d), respectively. The gaps in the light curve are due to the passage of the satellite through the South Atlantic Anomaly regions.}
\label{fig_2}
\end{figure}

\begin{table}
\small{
\caption{LAXPC count rates due to source (4U 0115+63) with background, background and source (background subtracted) in 3 energy bands.} 
\label{table2.tab}
\begin{tabular}{ccccccc}
\hline\hline
Observations&LAXPC  	& Time of Observation  	&Useful    &Energy	&Average  \\
            &           &                      	&Exposure  &Range	&Count\\
            &           &        	        &Time	   &		&Rate\\
            &           & (UT)                 	&(s)       &(keV)	&(counts/s)\\
\hline
Source+	  & 10&2015-10-24 09:17:02-16:08:46&16480&3-20&1060.1$\pm$0.25\\
Background	       				      &&&&20-40&80.9$\pm$0.07\\
		      				      &&&&40-80&96.6$\pm$0.08\\
	  & 20&2015-10-24 09:17:02-16:08:46&16479&3-20&1015.3$\pm$0.24\\
                                                      &&&&20-40&76.2$\pm$0.07\\
         					      &&&&40-80&76.6$\pm$0.07\\
	  & 30&2015-10-24 09:17:02-16:08:46&16483&3-20&1034.9$\pm$0.25\\
                                                      &&&&20-40&67.0$\pm$0.06\\
         					      &&&&40-80&98.0$\pm$0.08\\

Background& 10&2015-10-22 22:19:24.5 - 2015-10-23 02:01:42.03&28712&3-20&18.9$\pm$0.04\\
                                                      &&&&20-40&15.5$\pm$0.03\\
         					      &&&&40-80&71.1$\pm$0.09\\
	  & 20&2015-10-22 22:19:24.5 - 2015-10-23 02:01:42.03&28712&3-20&10.2$\pm$0.03\\
                                                      &&&&20-40&12.6$\pm$0.03\\
         					      &&&&40-80&50.6$\pm$0.06\\
	  & 30&2015-10-22 22:19:24.5 - 2015-10-23 02:01:42.03&28712&3-20&9.8$\pm$0.03\\
                                                      &&&&20-40&13.3$\pm$0.03\\
         					      &&&&40-80&60.1$\pm$0.07\\

Source& 10&2015-10-24 09:17:02-16:08:46&16480&3-20&1041.7$\pm$0.26\\
(Background					      &&&&20-40&65.5$\pm$0.07\\
subtracted)   					      &&&&40-80&25.7$\pm$0.08\\
	  & 20&2015-10-24 09:17:02-16:08:46&16479&3-20&1005.0$\pm$0.24\\
                                                      &&&&20-40&63.6$\pm$0.07\\
         					      &&&&40-80&26.0$\pm$0.07\\
	  & 30&2015-10-24 09:17:02-16:08:46&16483&3-20&1025.1$\pm$0.25\\
                                                      &&&&20-40&53.7$\pm$0.06\\
         					      &&&&40-80&37.8$\pm$0.08\\
\hline
\end{tabular}
}
\end{table}

\section{Data Analysis \& Results}

We have plotted Swift/BAT hard X-ray transient monitor light curve \citep{2013ApJS..209...14K} in Figure \ref{fig_1} starting from MJD 57300.0 for 2015 observation of 4U 0115+63\footnote{http://swift.gsfc.nasa.gov/results/transients/BAT\_detected.html}. The LAXPC observations of 2015 October 24 are indicated in Figure \ref{fig_1} by an open circle. The two NuSTAR observations are shown in the same figure by open squares. The LAXPC observed the source two days after the peak intensity of the outburst on 2015 October 22. 
\begin{figure}
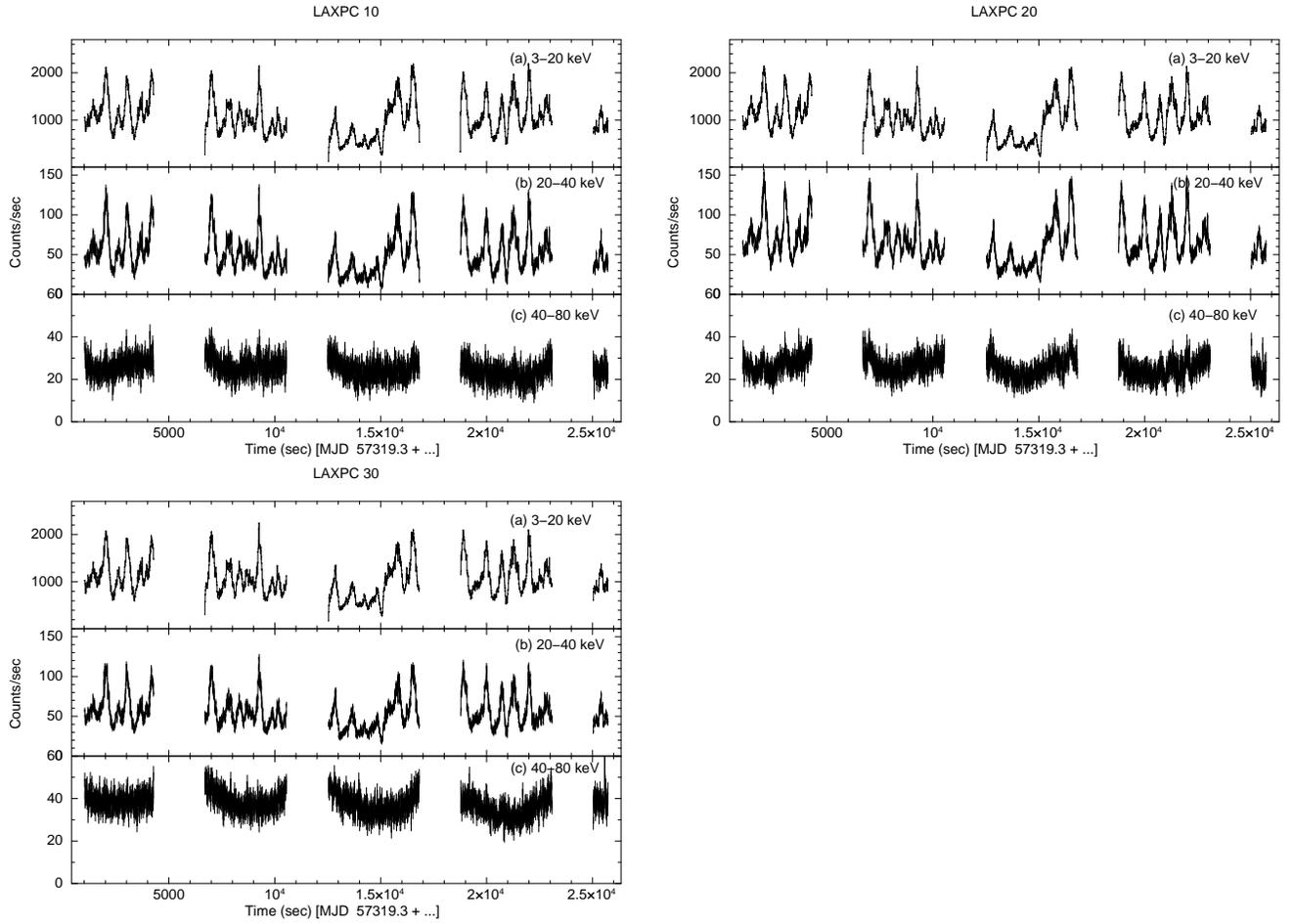

\includegraphics[scale=0.35,angle=-90]{fig3a.eps}
\includegraphics[scale=0.35,angle=-90]{fig3b.eps}
\includegraphics[scale=0.35,angle=-90]{fig3c.eps}
\caption{Background subtracted light curves are shown in panels (a) for 3-20 keV (b) 20-40 keV and (c) 40-80 keV, for LAXPC 10, LAXPC 20 and LAXPC 30 respectively.}
\label{fig_3}
\end{figure}

\subsection{LAXPC and NuSTAR Timing Analysis \& Results}
The observation log of the source and background is presented in Table \ref{table1.tab}. X-ray light curves have been extracted from 4 orbits of data in 3-80 keV for each LAXPC using the task "lxplc". Counts from all the 5 layers of each LAXPC have been summed up to construct the Light curves of 4U 0115+63 in 3-80 keV band for each LAXPC and shown in Figure \ref{fig_2}. 
Prominent intensity oscillations with $\sim$ 1000 s period are clearly visible in all the light curves. NuSTAR observations on 2015 October 22, one day prior to LAXPC observations, and 2015 October 30, 6 days after the LAXPC observations, show similar $\sim$ 1000 sec and $\sim$ 600 sec oscillations in 3-79 keV light curves shown in Figure \ref{fig_2}.

 To investigate energy dependence of the intensity variations, light curves in 3-20, 20-40 and 40-80 keV bands have been extracted for LAXPC 10, 20 and 30. For generating 3-20 keV light curve, only Layer 1 (outputs 1 and 2) data have been used as most incident photons ($\sim$ 90\%) of $<$ 20 keV energy are absorbed in Layer 1. Similarly light curves in 20-40 keV have been constructed using summed data from Layers 1 and 2 (2 outputs from each) and those in 40-80 keV from all the 5 layers. The count rates of each of the 3 LAXPCs in the 3 energy bands are shown in Table \ref{table2.tab}. Note that the higher background rate in LAXPC 10 is due to disabling of one side Veto layer due to its malfunction before the launch.
It can be observed from Figure \ref{fig_3} that the amplitude of $\sim$ 1000 sec oscillations is higher in 3-20 and 20-40 keV bands compared to that in 40-80 keV. Similar energy dependence is observed from LAXPC 20 and LAXPC 30 (Figure \ref{fig_3}). 

\begin{figure}
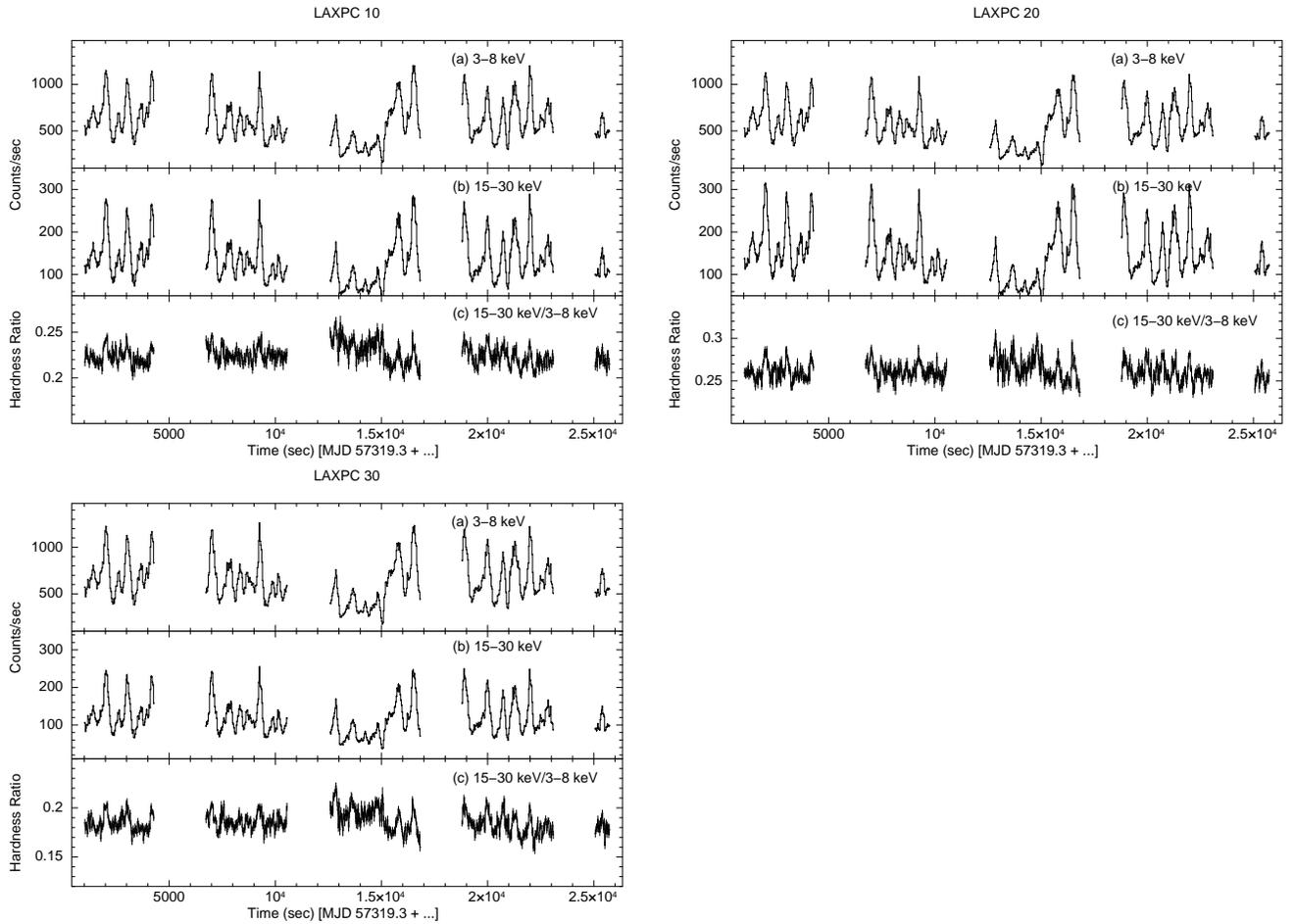

\includegraphics[scale=0.35,angle=-90]{fig4a.eps}
\includegraphics[scale=0.35,angle=-90]{fig4b.eps}
\includegraphics[scale=0.35,angle=-90]{fig4c.eps}
\caption{Background subtracted lightcurve for LAXPC 10, LAXPC 20 and LAXPC 30 in 3-8 keV band is shown in panel (a) and 15-30 keV band is shown in panel (b). Hardness ratio (HR) defined as ratio of counts in 15-30 keV/3-8 keV is plotted in panel (c) for LAXPC 10, LAXPC 20 and LAXPC 30.}
\label{fig_4}
\end{figure}

Hardness ratios (HR) defined as count rates in 15-30 keV/ 3-8 keV were derived to study spectral evolution during the intensity oscillations. These were computed using count rates from only top layer (output 1 and 2) for 3-8 and data from top 2 layers (outputs 1, 2, 3 and 4) in 15-30 keV. Selection of layers is made to obtain the optimal values of the hardness ratios. In Figure \ref{fig_4} background subtracted count rates in 3-8 and 15-30 keV in 30 sec bins are plotted along with HR as a function of time for LAXPC 10, LAXPC 20 \& LAXPC 30 to probe spectral changes during the oscillations. The intensity of the source shows no correlation with HR from all the 3 LAXPCs.
	
\begin{figure}
\includegraphics[scale=0.25,angle=-90]{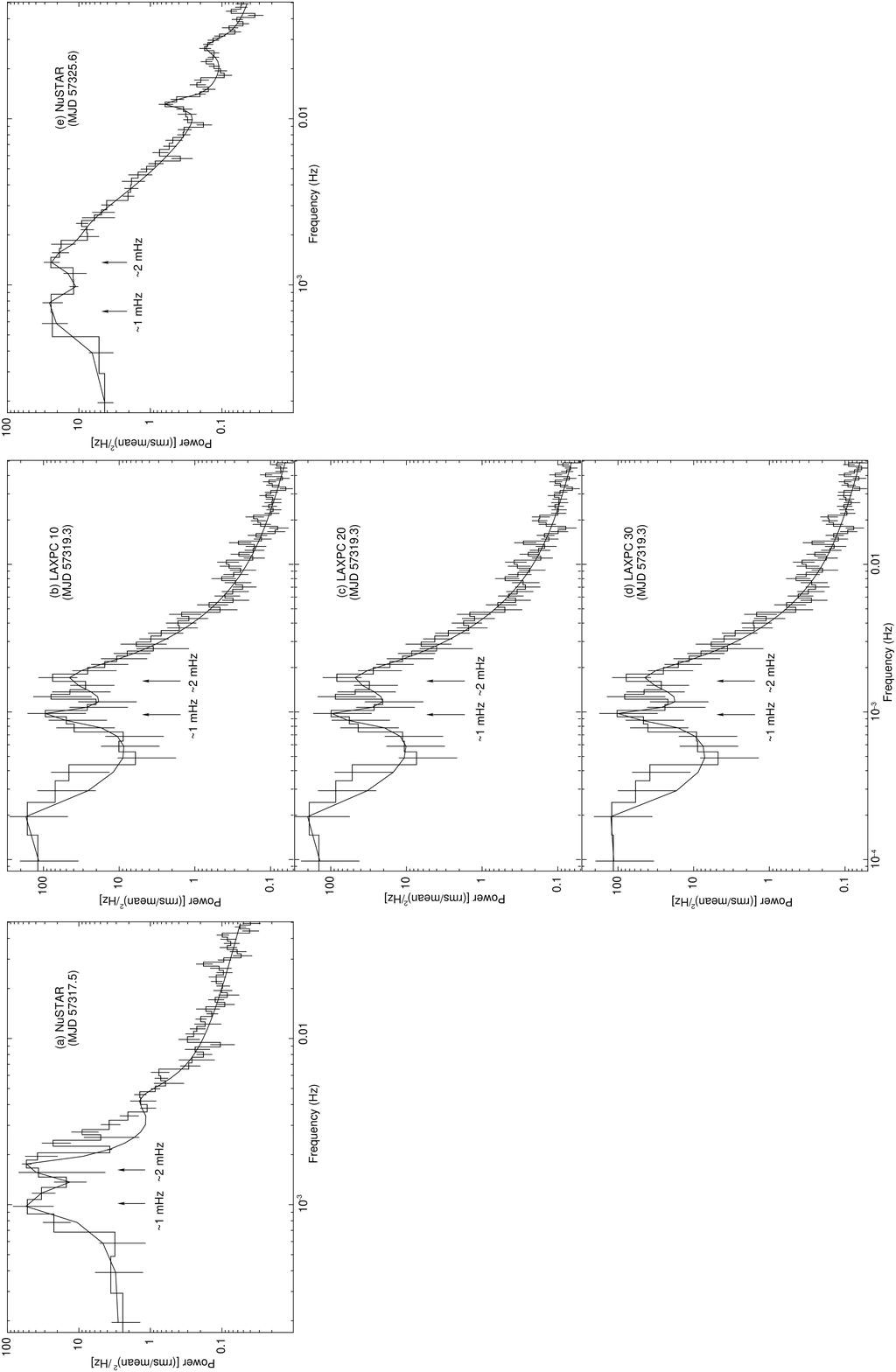}
\caption{The PDS (in the units of (rms/mean)$^{2}$/Hz) of 4U 0115+63 obtained for the outburst from NuSTAR in 3-79 keV energy band are shown in panels (a) and (e). PDS in 3-80 keV energy band of LAXPC 10, 20 and 30 are shown in panel (b), (c) and (d) respectively. Arrows indicates $\sim$ 1 mHz and $\sim$ 2 mHz QPOs in the PDS of NuSTAR and LAXPC detectors.}
\label{fig_5}
\end{figure}

PDS was generated using 10 seconds binned light curves in 3-80 keV for each of the 3 LAXPCs and 3-79 keV for the NuSTAR observations using ftool "powspec" in HEASOFT. The LAXPC light curves were divided into stretches of 1024 bins per interval. PDS from all the segments were averaged to produce the final PDS for the observation. Poissonian noise was subtracted from the PDS and they were normalized such that their integral gives the squared rms fractional variability normalized to units of (rms/mean)$^{2}$ /Hz. The PDS is fitted with two components, a power law and multiple Lorentzians to fit the QPO peaks. Figure \ref{fig_5} shows the PDS in which prominent peaks at $\sim$ 1 mHz (1000 sec) and $\sim$ 2 mHz (600 sec) are seen in all the three LAXPC spectra. We generated PDS for NuSTAR  using 10 sec binned light curves using the same procedure as used for generating the LAXPC PDS. NuSTAR observations of 2015 october 22 shows a single broad peak at $\sim$ 1 mHz with 1024
bins per interval whereas two clearly resolved peaks at $\sim$ 1 and 2 mHz are visible in the PDS with 512 bins per time interval. In NuSTAR PDS of 2015 october 30 two QPO peaks at  $\sim$ 1 and 2  mHz  are detected using 1024 as well as  512 bins per interval. We have, therefore, used 512 bins per interval for both the NuSTAR observations to generate PDS shown in Figure 5. The NuSTAR PDS of 2015, October 30 also shows presence of 12 mHz and 27 mHz QPOs  with coherence factor and RMS (\%) of 8 and 3 \% and 4 and 3\% respectively.
Summary of the characteristics of the $\sim$ 1 mHz and $\sim$ 2 mHz QPOs  from LAXPC and NuSTAR are presented in Table \ref{table3.tab}

To investigate energy dependence of the QPO characteristics, PDS were extracted  in 3-15, 15-30, 30-50 and 50-80 keV bands for each of the LAXPC units. This was achieved by extracting data from Layer 1 for 3-15 keV, Layers 1 \& 2 for 15-30 keV and Layers 1-5 data for 30-50 \& 50-80 keV energy bands. PDS in the units of (rms/mean)$^{2}$ /Hz are generated with 10 seconds binning and by subtracting Poissonian noise. Prominent QPOs are observed at $\sim$ 1mHz and $\sim$ 2 mHz in the energy bands up to 50 keV. Above 50 keV the 1 mHz and 2 mHz QPOs are insignificant. Details of QPO characteristics like frequency, RMS amplitude and quality factor are summarised in Table \ref{table3.tab}. The coherence factor ($\nu$/$\delta$$\nu$) lies in range 7.6-11.1 for $\sim$ 1 mHz QPO and 2.8-6.0 for the $\sim$ 2 mHz QPO respectively. It can be observed from Table \ref{table3.tab} that the coherence factor of $\sim$ 1 mHz QPO decreases from 3-15 keV to 15-30 keV energy band and then increases at higher energies. RMS amplitude (\%) of the $\sim$ 1 mHz QPO increases from 3-15 keV to 15-30 keV and then decreases at higher energies. It may be noticed that the coherence factor increases with energy for $\sim$ 2 mHz QPO and RMS amplitude (\%) of the $\sim$ 2 mHz QPO increases till 30 keV and then decreases at higher energy.

\begin{table}
\small{
\caption{Summary of the characteristics of the $\sim$ 1 mHz and $\sim$ 2 mHz QPOs observed from LAXPC and NuSTAR.} 
\label{table3.tab}
\begin{tabular}{cccccccccc}
\hline\hline
Observation&Detector&Exposure&Energy&\multicolumn{3}{c}{1 mHz QPO}&\multicolumn{3}{c}{2 mHz QPO}\\
\cline {5-7}
\cline {8-10}
   && &Range&Frequency & RMS&Quality &Frequency  &RMS&Quality  \\
(MJD)   &&(sec)&(keV)  &($\times 10^{-4}$ Hz) & \%&Factor &($\times 10^{-3}$ Hz)&\% &Factor\\
\hline\\
57317.5& NuSTAR &8584 & 3-79  &$10.5  ^{+8.5}_{-1.3}  $&15.7  &5.7    &$1.7   ^{+0.03}_{-0.2} $&15.8  &12.5 \\

57319.3&LAXPC 10 &16480& 3-80      &$9.4   ^{+1.5}_{-2.3}  $&13.7  &8.8    &$1.7   ^{+0.2}_{-0.2}  $&20.1  &2.7 \\
	&&&3-15               &$9.4   ^{+1.3}_{-4.0}  $&9.4   &9.0    &$1.7   ^{+0.2}_{-0.2} $&21.1  &3.0\\
        &&&15-30               &$9.6   ^{+1.4}_{-1.4}  $&12.2  &7.7    &$1.7   ^{+0.2}_{-0.2} $&23.2  &4.3\\
        &&&30-50               &$9.7   ^{+1.6}_{-1.6}  $&5.2   &8.9    &$1.7   ^{+0.2}_{-0.2} $&9.8   &4.4\\

57319.3&LAXPC 20 &16479& 3-80      &$9.4   ^{+1.7}_{-1.4}  $&14.3  &9.9    &$1.7   ^{+0.2}_{-0.2}  $&21.0  &2.6 \\
	&&&3-15               &$9.4   ^{+1.2}_{-3.3}  $&9.8   &8.2    &$1.7   ^{+0.2}_{-0.2} $&21.0  &3.4\\
        &&&15-30               &$9.5   ^{+1.3}_{-1.3}  $&12.2  &7.6    &$1.7   ^{+0.2}_{-0.2} $&24.1  &4.1\\
        &&&30-50               &$9.6   ^{+1.3}_{-2.0}  $&7.3   &8.6    &$1.7   ^{+0.4}_{-0.3} $&14.0   &6.0\\ 

57319.3&LAXPC 30 &16483& 3-80      &$9.5   ^{+1.6}_{-3.8}  $&13.8  &10.1    &$1.7   ^{+0.2}_{-0.2}  $&19.9  &2.7  \\
	&&&3-15               &$9.4   ^{+1.4}_{-4.1}  $&9.6   &9.6    &$1.7   ^{+0.2}_{-0.2} $&20.7  &2.8\\
        &&&15-30               &$9.6   ^{+5.7}_{-9.1}  $&12.5  &7.9    &$1.7   ^{+0.8}_{-0.4} $&22.9  &4.9\\
        &&&30-50               &$10.0  ^{+1.0}_{-4.3}  $&5.8   &11.5    &$1.7   ^{+0.2}_{-0.2} $&9.3   &5.8\\
57325.6 &NuSTAR&14564 & 3-79      &$6.9   ^{+0.5}_{-1.4}  $&10.9   &5.0    &$1.4   ^{+0.1}_{-0.1}  $&13.8  &2.32 \\

\hline
\end{tabular}
}
\end{table}

\subsubsection{$\sim$1 mHz \& $\sim$ 2 mHz QPOs from RXTE / PCA observations}

\begin{figure}
\includegraphics[scale=0.6, angle=0]{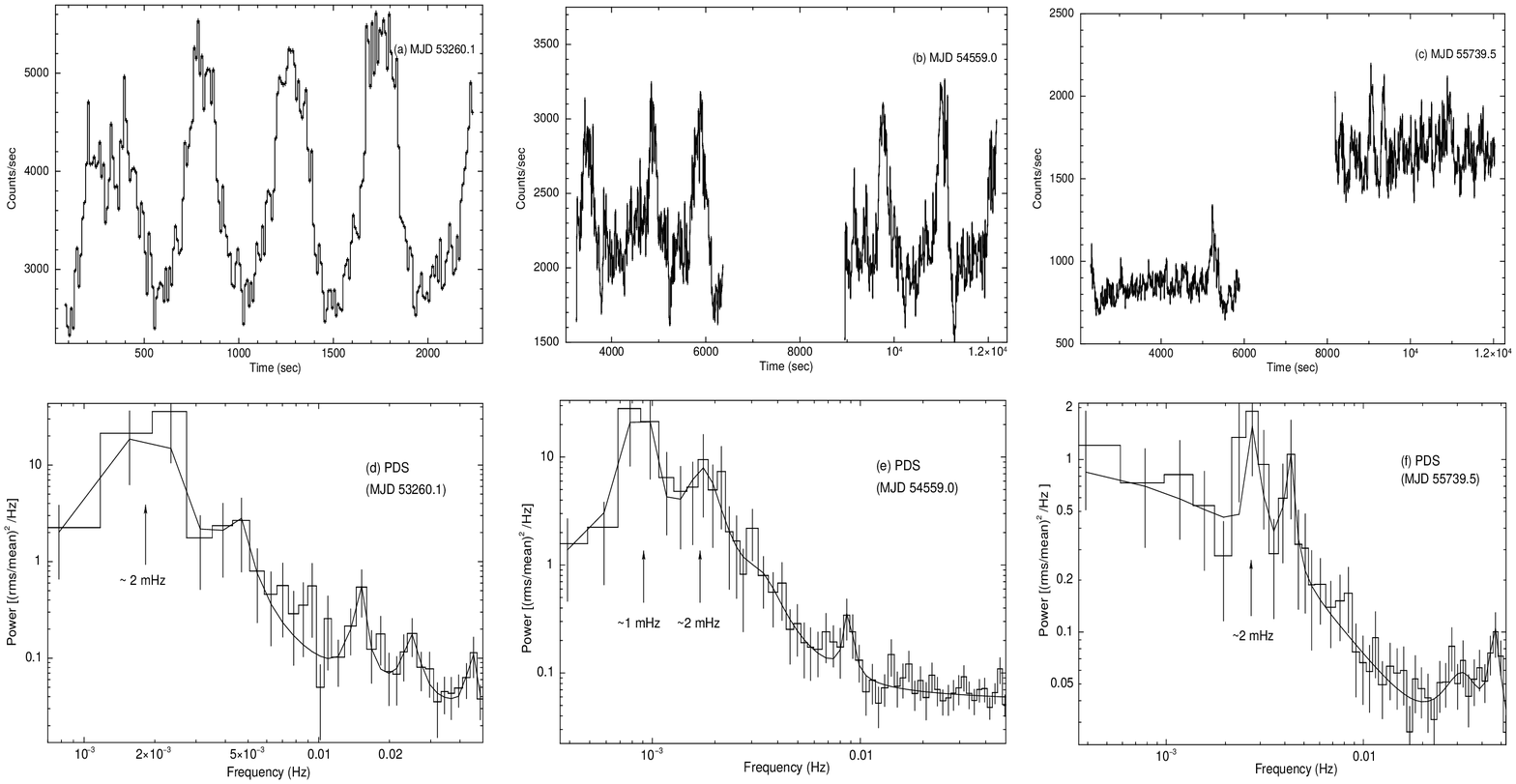}
\caption{Standard 1 mode light curves (10 seconds binned) in 2-60 keV energy band obtained from the RXTE/PCA observations from different outbursts: (a) 2004 September 12 (MJD 53260.1), (b) 2008 April 2 (MJD 54559.0) and (c) 2011 June 27 (MJD 55739.5) showing $\sim$ 1 mHz \& $\sim$ 2 mHz intensity oscillations are shown in the top panels. The change in the count rate seen in the light curve (c)  is due to one PCU being on during first half of the observation and 2 PCU being on for  the later part. PDS generated using 5 seconds binned light curves in 2-60 keV energy band from :  (d) 2004 September 12 (MJD 53260.1) showing $\sim$ 2 mHz intensity oscillation, (e) 2008 April 2 (MJD 54559.0) showing $\sim$ 1 mHz \& $\sim$ 2 mHz intensity oscillations and (f) 2011 June 27 (MJD 55739.5) showing $\sim$ 2 mHz intensity oscillation (indicated by arrows) of the corresponding lightcurve are shown in bottom panel.}
\label{fig_6}
\end{figure}

PCA/RXTE Standard 1 mode light curves (10 second binned) in 2-60 keV, generated using ftool "lcurve" in HEASOFT  for of all the observations showing $\sim$ mHz QPOs, are presented in Figure \ref{fig_6} (a), (b) and (c).  It may be noted that during 2004 observation, 3 PCUs were on while in 2008 observations 2 PCUs were on and the count rates in the figures reflect this. In the 2011 light curve in Fig 6 (c),  the jump in the count rate in the middle of the light curve is not due to intensity variation of the source but rather due to the fact that in the first half of the observation only one PCU was operational while in the later half two PCUs were operational. The observation of 2004 outburst on MJD 53260.1 is rather short (2159 sec)  seconds, and hence 5 sec binned light curve with 256 bins per interval was used for generating PDS using ftool "powspec" from this observation. Similarly 5 sec binned light curves were used to generate PDS for the 2008 and 2011 outbursts of the source. To resolve the mHz QPOs clearly and ensuring that at least two segments of light curves over which the PDS are averaged, we used 1024 and 512 bins per interval for the 2008 and 2011 observation respectively. Following procedures similar to that used for the LAXPCs, the PDS are fitted with two components, a power law and multiple Lorentzians to fit the QPO peaks from 7$\times$10$^{-3}$ - 0.05 Hz frequency range for 2004 observation and 3.5$\times$10$^{-3}$ - 0.05 Hz for 2008 and 2011 observations of the source. The resulting PDS in Figure \ref{fig_6} show prominent QPO peak in Figure \ref{fig_6}(d) \& Figure \ref{fig_6}(f) at $\sim$ 2 mHz (600 sec) and in Figure \ref{fig_6}(e) at $\sim$ 1mHz (1000 sec) and $\sim$ 2 mHz (600 sec) as indicated by arrows.
In Table \ref{table4.tab} we summarize the characteristics of mHz QPOs detected in our analysis of RXTE / PCA data during outbursts in 2004, 2008 and 2011 along with the 2 mHz QPO reported by \citet{1999ApJ...521L..49H} during the 1999 outburst. In addition to the 2 mHz QPO, three more  QPO peaks are detected in Figure \ref{fig_6}(d) at 14.9 mHz, 24.9 mHz and 44.7 mHz in the PDS of 2004 outburst. Coherence factor of the three QPOs are 11, 7 and 17 respectively. This result has been already reported by \citet{2013MNRAS.434.2458D}. They did not find any systematic variation between QPO frequency and flux for the 41 mHz QPO observed during different outbursts and 2mHz QPO observed during the same outburst. Similarly a QPO peak at 8.7 mHz of coherence factor 12, is visible in the PDS of 2008 outburst in Figure \ref{fig_6}(e) and a peak due to 46.4 mHz QPO of coherence factor 8, is visible in the PDS of 2011 outburst in Figure \ref{fig_6}(f).

\begin{table}
\small{
\caption{mHz QPOs detected in RXTE / PCA analysis of 4U 0115+63 along with those detected by LAXPC and NuSTAR.} 
\label{table4.tab}
\begin{tabular}{ccccccc}
\hline
Year of &Instrument&Observation Id&MJD& Detected QPOs (mHz) &Orbital  & Flux (3-50 keV)\\
outburst &&&&&Phase&$\times$10$^{-8}$ ergcm$^{-2}$s$^{-1}$\\
\hline\\
1999$^\dagger$          &RXTE (PCA)    &40411-01-09-00  &51248.3 &2   					 	&0.97&$1.81_{-0.01}^{0.02}$	\\
2004			&RXTE (PCA)    &90089-01-03-01  &53260.1 &1.9$^{+0.3}_{-0.6} $      			&0.70&$1.83_{-0.001}^{0.03}$	\\
                        &  	       &		&	 &(Q-factor=18.7, RMS=39.2\%) 			&    &				\\
2008			&RXTE (PCA)    &93032-01-03-03  &54559.0 &0.9$^{+0.2}_{-0.2} $,1.8$^{+0.5}_{-0.3} $	&0.12&$1.76_{-0.0004}^{0.02}$ 	\\
                        &  	       &		&	 &(Q-factor=8.2, RMS=14.5\%,                    &    &				\\
                        &  	       &		&	 &Q-factor=2.8, RMS=8.3\%)			&    &				\\
2011			&RXTE (PCA)    &96032-01-02-01  &55739.5 &2.9$^{+0.08}_{-0.14} $    			&0.66&$1.28_{-0.004}^{0.02}$ 	\\
                        &  	       &		&	 &(Q-factor=16.1, RMS=10.6\%) 			&    &				\\
2015			&NuSTAR        &90102016002     &57317.5 &1.1$^{+0.9}_{-0.1}$, 1.7$^{+0.03}_{-0.2}$	&0.56&$1.61_{-0.001}^{0.008}$	\\
2015			&LAXPC (10)    &9000000064  	&57319.3 &0.9$^{+0.2}_{-0.2}$, 1.7$^{+0.2}_{-0.2}$ 	&0.64&	-			\\
2015			&LAXPC (20)    &9000000064  	&57319.3 &0.9$^{+0.2}_{-0.1}$, 1.7$^{+0.2}_{-0.2}$  	&0.64&$1.23_{-0.014}^{0.106}$	\\
2015			&LAXPC (30)    &9000000064  	&57319.3 &0.9$^{+0.2}_{-0.4}$, 1.7$^{+0.2}_{-0.2}$  	&0.64&	-			\\
2015			&NuSTAR        &90102016004     &57325.6 &0.7$^{+0.05}_{-0.1}$, 1.4$^{+0.1}_{-0.1}$ 	&0.89&$1.09_{-0.0002}^{0.0168}$	\\
\hline\\
\end{tabular}
\flushleft{$^\dagger$Reported in \citet{1999ApJ...521L..49H}}
}
\end{table}

\subsection{Spectral Analysis \& Results}

We have performed detailed spectral analysis of only LAXPC 20 data as the spectral response matrix of this detector is well determined. Spectral analysis of LAXPC 20, NuSTAR \& RXTE data in 3-50 keV energy range was carried out for estimation of the source flux. Detailed spectral analysis of the complex spectrum of 4U 0115+63 will be described in a forthcoming paper. Data from all the 4 LAXPC orbits were merged to derive the source spectrum. The spectra were fitted with a model (``TBabs*powerlaw*highecut*cyclabs*cyclabs*cyclabs*cyclabs" inbuilt XSPEC), that included an interstellar absorption TBabs \citep{2000ApJ...542..914W} and combined model of powerlaw \citep{2007AstL...33..368T, 2013AstL...39..375B} and high energy cutoff \citep{1983ApJ...270..711W} 
\begin{equation}
f(E)=KE^{-\Gamma}\times \left\{ \begin{array}{ll}
1 & \mbox{($E \leq E_{c}$)}\\
exp^{-(E-E_{c})/E_{f}} & \mbox{($E > E_{c}$),} \end{array} \right.
\end{equation}
where, K is normalization factor (photons keV$^{-1}$cm$^{-2}$s$^{-1}$ at 1keV), $\Gamma$ is photon index of powerlaw, E is the photon energy, $E_{c}$ is the cut off energy in keV and $E_{f}$ is the e-folding energy in keV.
We also add cyclotron absorption line models (cyclabs XSPEC model, \citet{1990Natur.346..250M}) to account for the 4 known cyclotron lines at $\sim$ 11 to 15 keV, 23 keV, 35.6 keV, 48.8 keV (\citet{2009A&A...498..825F, 2006ApJ...646.1125N, 2013AstL...39..375B, 2015MNRAS.454..741I, 2002ApJ...580..394C}), defined as,
\begin{equation}
  M(E) =\exp\left[-D_{\rm f} \frac{(W_{\rm f} E/E_{\rm cycl})^{2}}{(E-E{\rm cycl})^{2}+W_{\rm f}^{2}}
 \right]
\label{equ03}
\end{equation}
Where D$_{\rm f}$ is depth of fundamental or first harmonic, E$_{\rm cycl}$ is the cyclotron energy, W$_{\rm f}$ is the width of fundamental or first harmonic. The CRSF line energies are independently determined by the fit. This was done to account for anharmonic line ratios which can occur due to effects of viewing geometry and asymmetric emission patterns \citep{2013PASJ...65...84N}.

The value of column density is fixed at, N$_{H}$=1.3$\times$ 10$^{22}$cm$^{-2}$ \citep{2015MNRAS.454..741I} for all the observations. The powerlaw photon index ($\Gamma$)=$0.35_{-0.06}^{0.04}$, highecut cutoff energy $E_{c}$=$8.18_{-0.22}^{0.25}$ keV and highecut fold energy $E_{f}$=$9.13_{-0.45}^{0.94}$ keV yielded the best fit with $\chi^{2}$/dof=86.68/59 for LAXPC 20  energy spectrum in 3-50 keV with 1\% systematic error. Flux estimated from this fit for LAXPC 20 is $1.23_{-0.01}^{0.1}$ $\times$ 10$^{-8}$ ergcm$^{-2}$s$^{-1}$ on 2015 October 24. Using the same model used for LAXPC,  with 1\% systematic error we derived NuSTAR flux to be $1.61_{-0.001}^{0.008}$ $\times$ 10$^{-8}$ ergcm$^{-2}$s$^{-1}$ on 2015 October 22 and $1.09_{-0.0002}^{0.02}$ $\times$ 10$^{-8}$ ergcm$^{-2}$s$^{-1}$ on 2015 October 30, before and after the LAXPC observation respectively. The Flux is obtained from the best fit using "flux" command in XSPEC and corresponding errors are estimated using "flux error" command in XSPEC in 3-50 keV energy range. The flux values are presented in Table \ref{table4.tab}. This flux is converted to luminosity to study the frequency of QPO dependence on X-ray luminosity presented in the next section. Detailed studies of the cyclotron lines and their characteristics along with the pulse phase resolved energy spectra, are under progress and will be reported separately.

\section{Discussion}
High frequency QPOs occur in $\sim$ 1 Hz to $\sim$ 1 kHz range quite commonly in Low mass X-ray Binaries (LMXBs) with a neutron star or a stellar-mass black hole as the X-ray  source. Low frequency QPOs of $\sim$ 10 to 100 mHz have been detected mostly in HMXB, being more common in transient pulsars in Be binaries. The reported QPO frequencies lie in 20-200 mHz as summarized by \citet{2010MNRAS.407..285J} with the exception of 4U~0115+63 in which \citet{1999ApJ...521L..49H} detected $\sim$ 2 mHz QPO during its 1999 outburst.

Recently \citet{2016MNRAS.460.3637S} detected a 1.46 mHz QPO in the HMXB IGR J19140+0951 from XMM-Newton observations during a flaring episode when the source flux increased more than 10 times. Larger radii of the accretion disks owing to high magnetic fields in HMXB systems, as compared to the black hole binaries or LMXB systems, will automatically cause any oscillations related to time-scales of the accretion disk to have lower frequencies in these systems. As described in the preceding sections, the HMXB 4U~0115+63 shows large amplitude and slow quasi periodic X-ray oscillations on timescale of 1000 sec and 600 secs. Although of a similar time scale as expected from HMXB systems, the large amplitude of the oscillations is possibly unique to this HMXB system. Similar kinds of periodic variabilities, albeit of shorter periods, have been observed from a few black hole X-ray binaries like GRS 1915+105 \citep{1998A&AS..128..145P} and IGR J17091-3624 in their $\rho$ variability class which show quasi periodic flares on the timescales of 40-120 seconds \citep{2011ApJ...742L..17A}. Millihertz oscillations have also been observed in dwarf nova systems (see \citet{2002ApJ...580..423M}) consistent with their larger accretion disk radii, but not with such large amplitudes.

Thus it would be natural to expect that the source of these oscillations in 4U~0115+63 has something to do with the accretion disk. The orbital phases for the reported QPOs in 4U 0115+63 from LAXPC, NUSTAR and RXTE observations are presented in Table \ref{table4.tab}. Orbital phase was calculated using periastron passage time (Tw) = 53243.038 of the 2004 outburst and an orbital period Porb = 24.3174 days \citep{2010MNRAS.406.2663R}.

\begin{figure}
\includegraphics[scale=0.8]{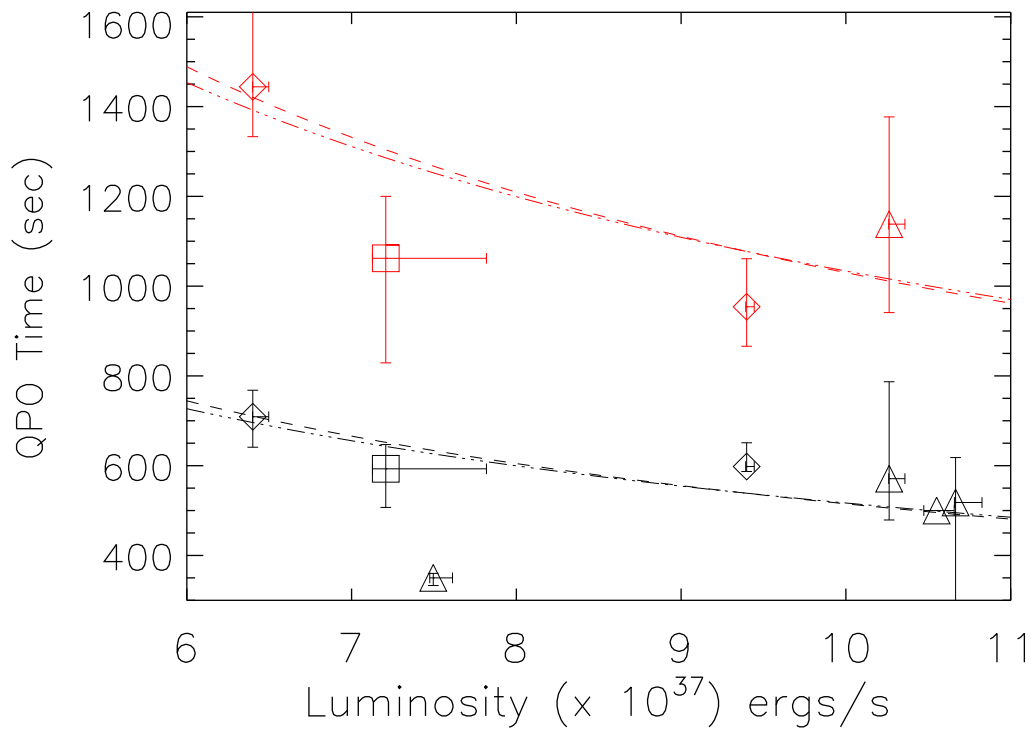}
\caption{Oscillation time period versus 3-50 keV X-ray luminosity, of 4U~0115+63. For each observation, the slowest time period oscillations ($\sim$ 1 mHz in red color) and its harmonic ($\sim$ 2 mHz in black color) have been plotted. The LAXPC observation is marked by a square box, NuSTAR with diamond and RXTE with triangle symbol. The dashed-dotted line corresponds to viscous time-scales (eqn. \ref{eqn:fin} ) and the dashed line corresponds to precession time-scales (eqn. \ref{eqn:prec} ). See text for details.}
\label{fig_7}
\end{figure}

Figure \ref{fig_7} shows a plot of the slowest observed QPO and its harmonic in the LAXPC, NuSTAR and PCA observations against their 3-50 keV luminosity. Since the LAXPC observation was performed close to the peak of the outburst (see Figure \ref{fig_1}) with a BAT count rate similar to that during the NuSTAR observation, we assume the QPOs from these three observations to be originating from a similar mechanism. The plot shows that there seems to be a trend in the observed oscillation frequency as a function of the luminosity of the source. Such a trend can then be used to constrain the models used to explain these oscillations.

The first possible explanation which can be put forth relates to oscillations tied to the keplerian frequencies at some fixed radii in the accretion disk. The keplerian frequency model (KFM) \citep{1987ApJ...316..411V} and the beat frequency model (BFM) \citep{1985Natur.316..239A, 1985BAAS...17R.860L} are popular models in this regard. For 4U~0115+63, with a magnetic field of $B{\star} \sim 10^{12} G$ as expected from its fundamental cyclotron line energy, the inner disk radius will correspond to its magnetospheric radius \citep{1979ApJ...232..259G, 2012A&A...544A.123B} of

\begin{equation}
  R_m = 273\, \left( \frac{\Lambda}{0.1} \right)
	  \left( \frac{M{\star}}{1.4~M_{\odot}} \right)^{1/7}
	  \left( \frac{R{\star}}{10\, km} \right)^{10/7}
	  \left( \frac{B{\star}}{10^{12}\,G} \right)^{4/7}
	  \left( \frac{Lx}{10^{37}\, erg\,s^{-1}} \right)^{-2/7}
	  \, km
   	\label{eqn:ralf}
\end{equation}

where, $M{\star}$ and $R{\star}$ correspond to mass and radius of the neutron star and $Lx$ the X-ray luminosity. $\Lambda$ for disk accretion is given as $\approx 0.22 \alpha^{18/69}$, with $\alpha$ being the disk viscosity parameter. The keplerian frequency $\nu_{KFM} = \frac{1}{2\pi}\sqrt{GM/R^3}$ at this radius would be $\sim 15 Hz$, much greater than the mHz time periods required for both the keplerian and beat frequencies. A second possibility, as suggested by \citet{1999ApJ...521L..49H} who reported 2 mHz oscillation, would involve obscuring / scattering matter at the outer edge of the accretion disk. Millihertz frequencies correspond to keplerian frequency at radii $\sim 10^{10}$ cm. For this binary system, with a projected orbital distance of the neutron star from its center of mass of $\sim 140 lt-sec \simeq 10^{12}$ cm, it seems unlikely that the outer disk radius would be so small. Thus, it would seem that both KFM and BFM cannot explain the observed millihertz oscillations.  

An explanation for the oscillations in GRS~1915+105 has used thermal disk instabilities occurring in the inner accretion disk, when the local mass accretion rate approaches its Eddington limit. The thermal instability results in large amplitude X-ray QPO in the viscous timescales of a few minutes to hours. A similar explanation involving viscous relaxation time-scales caused by the thermal disk instabilities can be used for the $\sim$ 1000 sec oscillations, as the source was at the peak of the outburst when these oscillations were observed (close to Eddington luminosity). Viscous time-scales for HMXB accreting source (\citet{2002apa..book.....F}, page 113, equation 5.69) are given as

  	\begin{equation}
	  t_{visc} = 480 \, \alpha^{-4/5} 
	  \left( \frac{\dot{M}}{10^{17}\, g/s} \right)^{-3/10}   
	  \left( \frac{M{\star}}{M_{\odot}} \right)^{1/4} 
	  \left( \frac{R_{m}}{1000\, km} \right)^{5/4}
	  \,  s
	  \label{eqn:tvisc}
	\end{equation}

Using $Lx \simeq G M \dot{M} / R{\star}	$, this gives 
	\begin{equation}
	  t_{visc} = 991.1 \, 
	  \left( \frac{Lx}{10^{37}\, erg \,s^{-1}} \right)^{-23/35} 
	  \left( \frac{\alpha}{0.1} \right)^{-109/230}
	  \left( \frac{R{\star}}{10\,km} \right)^{52/35}
	  \left( \frac{M{\star}}{1.4\, M_{\odot}} \right)^{-51/70}
	  \, s
	  \label{eqn:fin}
	\end{equation}

As seen from the dashed-dot line in Fig. \ref{fig_7}, this expression does follow the trend of oscillation time scales with luminosity.

However, \citet{2016Natur.529...54K} observed 100 seconds to 2.5 hours variability in optical and X-ray wave bands from the transient black hole binary V404 Cygni, similar to the X-ray timescales of GRS 1915+105. The optical variability was observed at a mass accretion rate which was 10 times lower than the Eddington limit. This shows that the reason for the disk instabilities leading to the viscous time-scale periodicities, need not be related to thermal instabilities at high accretion rates.

  Another possibility for causing such milliHz oscillations was suggested by \citet{2002ApJ...565.1134S}. They reasoned that a warped or precessing accretion disk caused by mis-aligned vectors of the disk angular momentum and the magnetic field can lead to such observational signatures. The frequency of warping / precession at the magnetospheric radius as given by eqn. 27 in their paper lies in the millihertz range with 

  \begin{equation}
	\nu_{qpo} \propto (0.83 mHz) \,
	\left( \frac{\dot{M}}{10^{17} \, g/s} \right)^{0.71} 
	\left( \frac{\alpha}{0.1} \right)^{0.85}
	\label{eqn:prec1}
  \end{equation}
  
  This can be re-cast to give an equation similar to \ref{eqn:fin}, giving

 \begin{equation}
	t_{prec} = 775.9 \,
	\left( \frac{{Lx}}{10^{37}\, erg \,s^{-1}} \right)^{-0.71}
	\left( \frac{\alpha}{0.1} \right)^{0.85}
	\label{eqn:prec}
  \end{equation}

  As seen in Figure \ref{fig_7}, both the models fit the data points well and hence it is difficult to distinguish between these two models. It is worth noting though, that the best fit $\alpha$ (0.023) value for the precession model is higher than the best fit value of $\alpha$ (0.016) for the viscous disk instability model. More detailed observations of mHz QPOs in 4U 0115+63 over a much wider range of luminosity are required to discriminate between the two models and determine origin of these low frequency QPOs with more certainty.

\section*{Conclusion}

We have detected $\sim$ 1 mHz and 2 mHz oscillations from observations with the LAXPC instrument onboard AstroSat during the peak of the 2015 outburst close to the periastron passage of the pulsar. Similar low frequency QPOs are also detected in the NuSTAR data obtained before and after the LAXPC observations. \citet{1999ApJ...521L..49H} had earlier reported the detection of $\sim$ 2 mHz QPO during the same phase of periastron passage in 1999 outburst. We have detected similar low frequency QPOs during the 2004, 2008 and 2011 outbursts from the RXTE/PCA data. This suggests the possibility of this phenomenon being related to the elliptical binary orbit of the Be/X-ray pulsar. Present detection of $\sim$ 1 mHz \& 2 mHz oscillations from the X-ray pulsar 4U 0115+63 implies that the quasi periodicity timescale is independent of the accretor and depends on accretion and physical parameters of the binary system which is distinct from other smaller time scale QPOs observed from this and other binary systems. It is of utmost importance to model the different physical origins of slow and fast quasi periodicities in a unified way irrespective of the nature of the accretor.

\section*{Acknowledgements}
We thank members of the LAXPC team for their contribution to the development of the LAXPC instrument. We also acknowledge contributions of the AstroSat project team at ISAC. This research has made use of data obtained through the HEASARC Online Service, provided by the NASA/GSFC, in support of NASA High Energy Astrophysics Programs. JR and PCA acknowledge the fellowship and the funding provided by the National Academy of Sciences, India (NASI). This paper makes use of data from the AstroSat mission of the Indian Space Research Organisation (ISRO), archived at the Indian Space Science Data Centre (ISSDC). Finally we thank the anonymous referee for critical observations and comments that considerably improved the content of the paper.


\bibliographystyle{apj}

\end{document}